\documentclass[conference]{IEEEtran}
\IEEEoverridecommandlockouts
\usepackage{cite}
\usepackage{amsmath,amssymb,amsfonts}
\usepackage{algorithmic}
\usepackage{graphicx}
\usepackage{textcomp}
\usepackage{xcolor}
\usepackage{lipsum}
\usepackage{caption}
\usepackage{subcaption}
\def\BibTeX{{\rm B\kern-.05em{\sc i\kern-.025em b}\kern-.08em
    T\kern-.1667em\lower.7ex\hbox{E}\kern-.125emX}}
\usepackage{placeins}
\usepackage{tabularx}
\usepackage{longtable}
\usepackage{fontawesome}
\usepackage{soul}
\usepackage{tcolorbox}
\usepackage{colortbl}
\usepackage{changepage}
\usepackage{framed,enumitem}
\usepackage{setspace}
\usepackage{hyperref}

\usepackage{multirow}
\usepackage{orcidlink}
\usepackage{rotating}
\usepackage{environ}
\usepackage{pdflscape}% not required
\usepackage{stmaryrd}
\newcommand{\interviewquote}[2]{
 \def\FrameCommand{%
    \hspace{0pt}%
    {\color{blue}\vrule width 1.5pt}% 
    {\color{white}\vrule width 4pt}%
    \colorbox{white}
  }%
  \MakeFramed{\advance\hsize-\width\FrameRestore}%
  \noindent\hspace{-4.55pt}% disable indenting first paragraph
  \begin{adjustwidth}{}{7pt}
  {\footnotesize ``\emph{#1}'' - {#2}}\vspace{0.1pt}\end{adjustwidth}\endMakeFramed%
}
\usepackage{ifthen}
\usepackage{amssymb}
\newboolean{showcomments}
\setboolean{showcomments}{true} % toggle to show or hide comments and proof reading markings

\ifthenelse{\boolean{showcomments}}
  {\newcommand{\nb}[2]{
    \fcolorbox{gray}{yellow}{\bfseries\sffamily\scriptsize#1}
    {\sf\small$\blacktriangleright$\textit{#2}$\blacktriangleleft$}
   }
   
  }
  {\newcommand{\nb}[2]{}
   
  }

\usepackage[normalem]{ulem} % for \sout
\usepackage{xcolor}

\ifthenelse{\boolean{showcomments}}
  {
  
  \newcommand{\del}[1]{\textcolor{red}{\sout{#1}}} % please delete
  \newcommand{\inspar}[1]{\color{blue}{#1}\color{blue}} % Put edit comments in a really ugly standout display
  }
  {
  
  \newcommand{\del}[1]{} % Removes text
  \newcommand{\inspar}[1]{} % Removes edit and box
  
  }

\begin{document}

%\title{On data, annotations, and ecosystem challenges of data-intensive software development in the automotive industry}
\title{Automotive Perception Software Development: An Empirical Investigation into Data, Annotation, and Ecosystem Challenges}
%\alessia{I think the title is a bit too broad to reflect the content of the paper. My suggestion would be: Data, Annotations, and Ecosystem Needs in AI-based Software Development. We could add also the automotive industry if it is not too long. Why do we have the * by the way in the title?}
%\martin{I modified the title. I would like to keep automotive industry because SEIP has a focus on software in practices, i.e., industry. Is the English correct in the title? I'm unsure if it is of data-intensive software or in data-intensive software}

\author{
\IEEEauthorblockN{Hans-Martin Heyn\IEEEauthorrefmark{1}\IEEEauthorrefmark{4},
Khan Mohammad Habibullah\IEEEauthorrefmark{1},
Eric Knauss\IEEEauthorrefmark{1},
Jennifer Horkoff\IEEEauthorrefmark{1}
\\
Markus Borg\IEEEauthorrefmark{2},
Alessia Knauss\IEEEauthorrefmark{3},
Polly Jing Li\IEEEauthorrefmark{5}}

\IEEEauthorblockA{\IEEEauthorrefmark{1}\textit{Chalmers $\mid$ University of Gothenburg, Sweden}\\
\IEEEauthorrefmark{2}\textit{RISE Research Institutes of Sweden} \\
\IEEEauthorrefmark{3}\textit{Zenseact AB, Sweden},
\IEEEauthorrefmark{5}\textit{Kognic AB, Sweden} \\
}
\IEEEauthorblockA{\IEEEauthorrefmark{4}Corresponding author, Hans-Martin.Heyn@gu.se}
}

\maketitle

%%%%%%%%%%%%%%%%%%%%%%%%%%%%%%%%%%%%%%%%%%%%%%%%%%%%%%%%%%%%%%%%%%%%%%%%

\begin{abstract}
Software that contains machine learning algorithms is an integral part of automotive perception, for example, in driving automation systems. 
The development of such software, specifically the training and validation of the machine learning components, require large annotated datasets. 
An industry of data and annotation services has emerged to serve the development of such data-intensive automotive software components. 
Wide-spread difficulties to specify data and annotation needs challenge collaborations between OEMs (Original Equipment Manufacturers) and their suppliers of software components, data, and annotations.\par
This paper investigates the reasons for these difficulties for practitioners in the Swedish automotive industry to arrive at clear specifications for data and annotations. The results from an interview study show that a lack of effective metrics for data quality aspects, ambiguities in the way of working, unclear definitions of annotation quality, and deficits in the business ecosystems are causes for the difficulty in deriving the specifications. We provide a list of recommendations that can mitigate challenges when deriving  specifications and we propose future research opportunities to overcome these challenges. 
Our work contributes towards the on-going research on accountability of machine learning as applied to complex software systems, especially for high-stake applications such as automated driving.
\end{abstract}

%%%%%%%%%%%%%%%%%%%%%%%%%%%%%%%%%%%%%%%%%%%%%%%%%%%%%%%%%%%%%%%%%%%%%%%%


%%%%%%%%%%%%%%%%%%%%%%%%%%%%%%%%%%%%%%%%%%%%%%%%%%%%%%%%%%%%%%%%%%%%%%%%
%%%%%%%%%%%%%%%%%%%%%%%%%%%%%%%%%%%%%%%%%%%%%%%%%%%%%%%%%%%%%%%%%%%%%%%%
%%%%%%%%%%%%%%%%%%%%%%%%%%%%%%%%%%%%%%%%%%%%%%%%%%%%%%%%%%%%%%%%%%%%%%%%

\section{Introduction}
Driving automation refers to system that can automatically intervene in the driving task \cite{SAEJ3016}. This includes advanced driver assistance systems (ADAS) which can be seen as a pre-stage to conditional or even full autonomous driving \cite{Kukkala2018}. 
The aim of ADAS is to provide comfort and especially additional safety to manual driving tasks because the majority of accidents are still caused by human error \cite{Khattak2021}. 
Prominent features of ADAS include, among others, \emph{collision avoidance}, \emph{lane detection}, \emph{traffic sign recognition}, \emph{pedestrian detection}, \emph{parking assistance} and \emph{driver monitoring} \cite{Chacon2015}. 
All these features rely on the availability of data from a great variety of sensors, e.g., cameras, radar, lidar, ultrasonic sensors, fused and processed in real-time in what is referred to as the \emph{perception system}. 
Besides rule-based software components, the automotive industry relies on machine learning (ML) to enable the perception system of ADAS to be fast, robust, and precise enough in processing the incoming data\cite{Moujahid2018}. 
The advent of ML in the automotive industry, however, has caused a paradigm shift, because software engineers no longer express all logic in source code. Instead, they train ML models with large, often pre-annotated datasets. Traditional processes for specifying, developing, and testing automotive software can no longer apply. Correctly working software that incorporates ML algorithms require not only the avoidance of systematic mistakes during software development but also sets expectations on the datasets available at design- and run-time \cite{Borg2019}, including the annotation of such data.

\begin{figure}[!b]
    \centering
    \includegraphics[width=\linewidth]{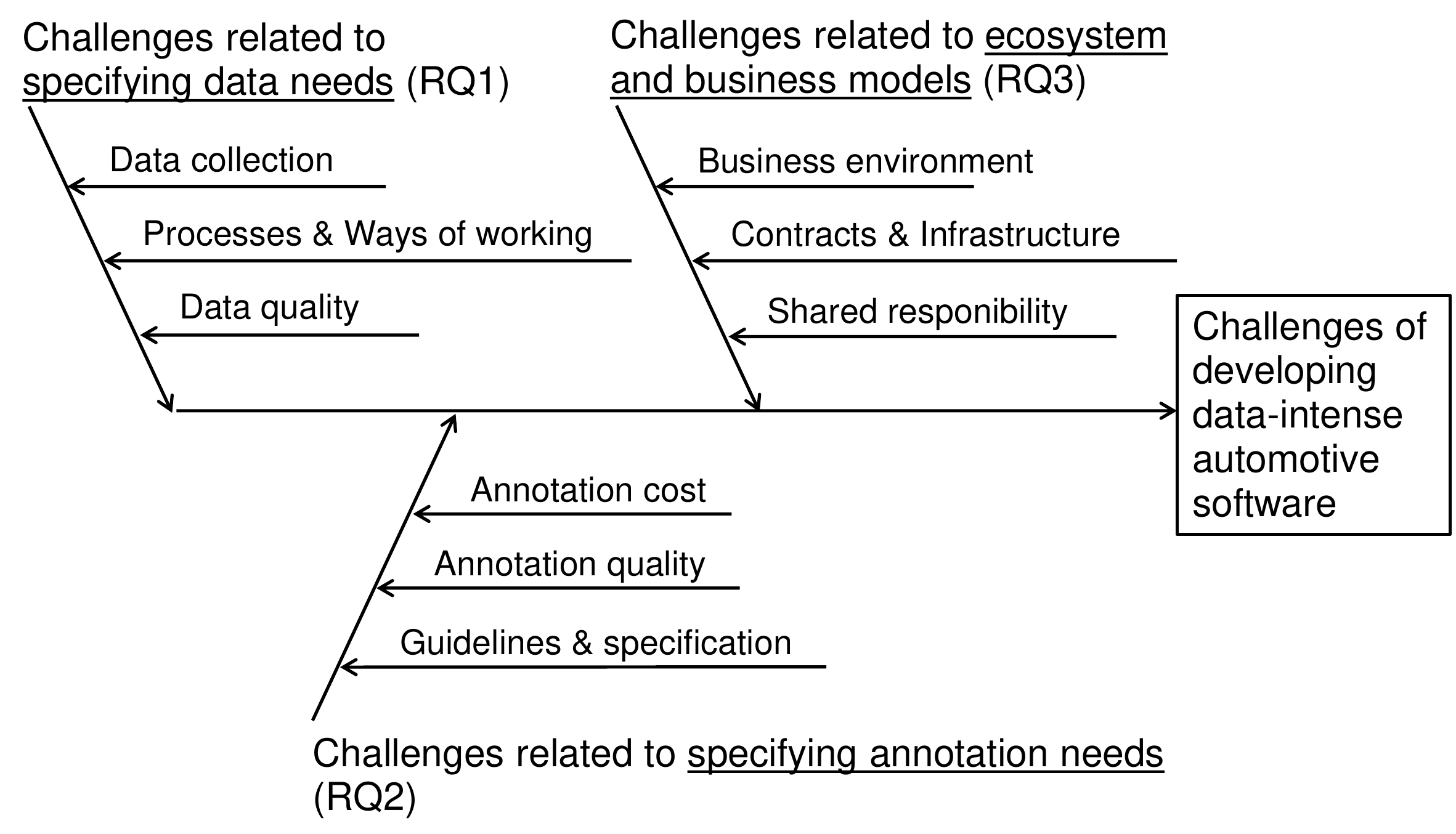}
    \caption{Cause-Effect diagram showing the major themes regarding the ability to specify data and annotation needs, and the ability of ecosystems and business models to handle shared data-intensive software development.}
    \label{fig:cause_effect_all}
\end{figure}

%\del{In summary, ADAS and subsequently autonomous driving (AD) requires successful perception, which relies on data driven software development that includes ML.}
%The performance of the software depends on the performance of the integrated ML models which, in turn, depend on the quality of data used during training and encountered at runtime. 
The datasets must fulfil the desired expectations in order to fulfil the desired performance of the software.
%Machine learning builds on high-quality data. High-quality data means that the datasets fulfil the desired expectations in regards to the later performance of the ML model. 
In our pre-study \emph{Precog} we explored how practitioners specify expectations on ML models, data, and data annotations as well as which trade-offs are taken for different quality aspects in software serving automotive perception systems, see \cite{Precog2022} for more detailed information. 
Through an interview-based study with industrial practitioners from the Swedish automotive industry we identified challenges in eight themes: \emph{AI and ML models}, \emph{annotation}, \emph{data}, \emph{ecosystem and business}, \emph{quality}, \emph{requirement engineering}, \emph{perception}, and \emph{system and software engineering}.
In this article, we investigate closer the challenges identified in the data, annotation, and ecosystem and business themes. The reason we combine these themes together is the current advent of a \emph{data industry} in the automotive supplier sector \cite{Salinesi2018,Martin2020,Wang2022}. Original Equipment Manufacturers (OEMs) rely heavily on the collaboration with suppliers for the development of vehicle systems and software. The success of such collaborations depend on the ability of OEMs to specify expectations towards its suppliers. 
For the shared development of data-intensive software, this implies that both the OEM and the supplier must be able to specify and understand expectations on data, expectations on the annotation of data, and to maintain suitable interactions and collaborations for handling data-intensive software development.\par
%In this study, we identified challenges specifying the data and annotation needs for the development of software that includes machine learning components in automotive perception systems as an important problem. 
%This study identifies reasons for the lack of data and annotation specifications, and it illuminates \alessia{we can iterate a bit more on this text. We do not really elicit any business models capabilities, maybe better to use investigate? We used ecosystem needs at some point in the paper, I believe that could be a nice formulation for this part of the text.} current practices of handling data-intensive development necessary for automotive perception systems in relation to software development ecosystems and business models in the automotive industry. \par
The study is guided by three research questions:
\begin{description}
\item[RQ1] What challenges do practitioners experience when specifying data needed for the development of automotive perception software that include machine learning components?
\item[RQ2] What challenges do the same practitioners experience when specifying annotation needs for data as part of the software development process?
\item[RQ3] What are implications towards industry ecosystems and business models for handling shared development of data-intensive software?
\end{description}
Figure~\ref{fig:cause_effect_all} gives an overview of the main themes we found in relation to the research questions. 
Concerning RQ1, we found that the ability to specify data needs is negatively impacted by nontransparent data selection as part of the data collection process, missing process guidelines and a lack of common metrics describing data variation as a means of representing data quality. In regards to RQ2, the most critical challenges we found are inconsistent manual annotations and missing specifications and guidelines for the annotation processes. In answering RQ3, we found that, in relation to the business environment, conventional value chains and sourcing policies impede shared data-intensive software developments. We saw a trend towards sharing development tools and utilising open source policies. Furthermore, we saw that new forms of collaborated development and contracts are required to facilitate transparency and shared responsibility in data driven developments.

\section{Related Work} \label{sec:relatedwork}
Rahimi et al. called for more attention from the requirement engineering community towards the ability of specifying of, what they referred to as, \emph{Machine-Learned Components} (MLC) \cite{Rahimi2019}. They explicitly mentioned datasets as an own aspect of MLCs that need to be properly specified for achieving a desired outcome of the MLC. Especially in safety-critical perception systems, requirements need to be specified towards robustness of the MLC \cite{Hu2020}. Robustness is achieved if \emph{small} changes in the input images do not lead to undesired behaviour. However, often it is not clearly specified what \emph{small} changes in the input space entail \cite{Hu2022}. In an interview study with data scientists, Vogelsang and Borg identified gaps in mutual understanding of technical concepts and measures between customers and data scientists who prepare data for ML models as a consequence of the lack of proper data specifications \cite{Vogelsang2019}. The problem of missing data specifications becomes apparent if software application-specific requirements are to be incorporate as prior domain knowledge into the training dataset \cite{Zhou2017}. Another consequence of a lack of context-based specification for datasets is that the datasets tend to become intractably large which makes it impossible to scrutinise their content \cite{Bender2021}. \par
Paullada et al. elaborate on several challenges encountered in data handling in ML research that also seem to apply towards commercial software application of ML \cite{Paullada2021}. Jo and Gebru, for example, argue that unspecified data collection resembles a \emph{wild west} mentality resulting in the risk of bias in the datasets. Instead, they propose the use of documentation methods from archiving, such as \emph{mission statements} and \emph{process records} \cite{Jo2020}. Similar to model cards for ML \cite{Mitchell2019}, datasheets for data have been proposed as a first step towards data specifications \cite{Gebru2021}.\par 
Besides a lack of data specifications, there is also a lack of specifications for the annotations of the data. Most commonly, data for ML training is annotated manually, sometimes even with the help of \emph{crowd-working} platforms. The use of crowd-working however can obscure the annotation process \cite{Hube2019}. A clear task design is required to avoid human-induced errors which again calls for a proper requirements specification \cite{Vaughan2017}. Even the definition of success metrics impact the result of the annotation process: for example a high annotation \emph{accuracy} not necessarily results in high correctness of model predictions \cite{Tsipras2020}. In automotive use cases of ML, the under-specification of data and annotations causes ambiguities in the requirements towards data which has negative implications towards the verifiability of safety-critical software that uses ML \cite{Salay2018, Borg2019}. For example, internal audits that should ensure correct behaviour of a ML model in relation to a company's ethical values cannot be conducted without proper specification of the data collection and processing \cite{Raji2020}. Accountability can only be established through transparency and ownership of the dataset development lifecycle which requires rigorous documentation of each stage in it \cite{Hutchinson2021}. Initial attempts to create large publicly available datasets are for example nuScenes for autonomous driving \cite{nuscenes2020}. \par
%Software development ecosystems have been formed to accommodate software development across different companies.
Knowledge sharing, quality definition, and effective communication are some necessary aspect of running a shared software development ecosystems \cite{Alves2017}. However, in cross-company software development projects, a lack in documented domain assumptions and missing cross-organisational documentation, e.g., data specifications, during data collection, labelling, and cleaning have been identified as a major cause for failures of the resulting software that contains ML models \cite{Sambasivan2021}.

%%%%%%%%%%%%%%%%%%%%%%%%%%%%%%%%%%%%%%%%%%%%%%%%%%%%%%%%%%%%%%%%%%%%%%%%
%%%%%%%%%%%%%%%%%%%%%%%%%%%%%%%%%%%%%%%%%%%%%%%%%%%%%%%%%%%%%%%%%%%%%%%%
%%%%%%%%%%%%%%%%%%%%%%%%%%%%%%%%%%%%%%%%%%%%%%%%%%%%%%%%%%%%%%%%%%%%%%%%

%\section{Problem statement} \label{sec:problem}

%%%%%%%%%%%%%%%%%%%%%%%%%%%%%%%%%%%%%%%%%%%%%%%%%%%%%%%%%%%%%%%%%%%%%%%%
%%%%%%%%%%%%%%%%%%%%%%%%%%%%%%%%%%%%%%%%%%%%%%%%%%%%%%%%%%%%%%%%%%%%%%%%
%%%%%%%%%%%%%%%%%%%%%%%%%%%%%%%%%%%%%%%%%%%%%%%%%%%%%%%%%%%%%%%%%%%%%%%%

\section{Method} \label{sec:method}
The study bases on a series of group interviews and a workshop conducted with different stakeholders involved in the development of perception systems for driving automation in the Swedish automotive industry. 
The reasons for choosing the Swedish automotive industry was on the one hand convenience of having local access to the stakeholders, and on the other hand the high competitiveness and international interdependence of the Swedish automotive industry.\par 
With some Swedish OEMs being subsidiaries of larger international companies, such as Geely or Volkswagen AG, or being themselves large international players with several subsidiaries, such as Volvo Group AB, the Swedish automotive industry provides in our opinion a representative sample of the worldwide automotive industry.\par
\begin{figure}[!htbp]
    \centering
    \includegraphics[width=\linewidth]{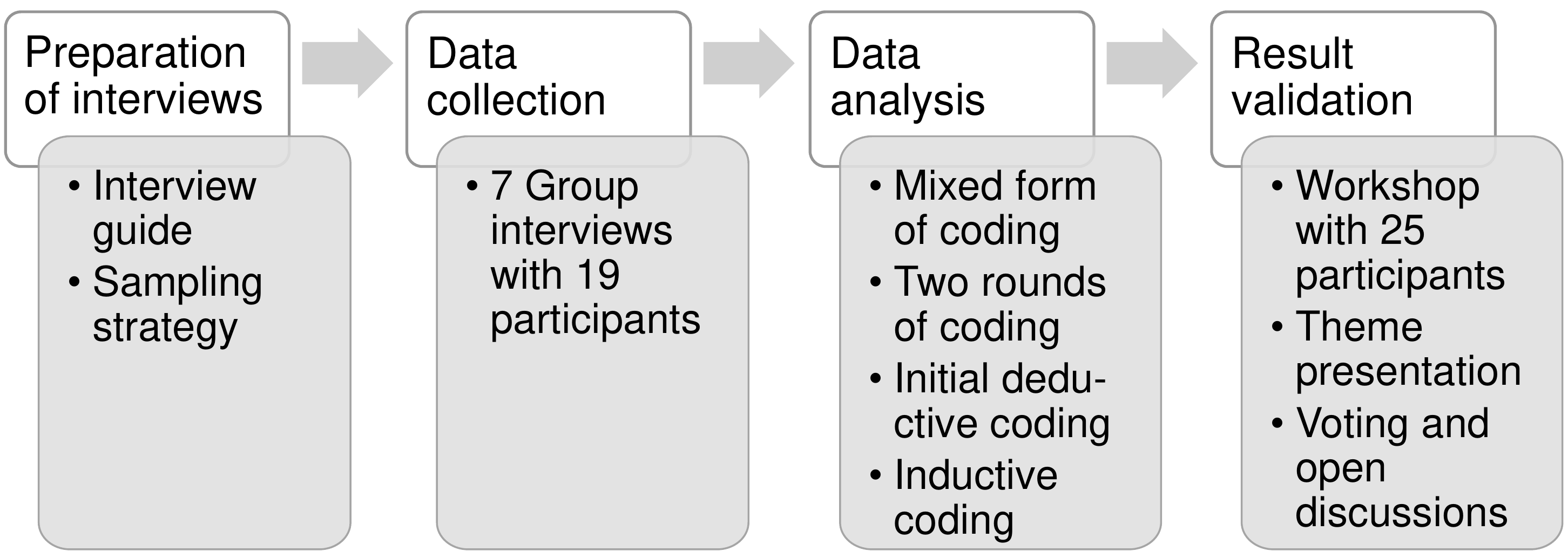}
    \caption{Overview of the interview study}
    \label{fig:method}
    \vspace{-1em}
\end{figure}

Figure~\ref{fig:method} depicts the research process followed during the study. 
%\alessia{Does the 25 participant workshop in data collection refer to the same thing as the 2.5 hour workshop in the validation? Maybe it would be good to split the interviews for deriving the themes and the validation workshop into two parts of the study, which both have preparation, data collection and analysis, but the interview analysis would be used as preparation for the workshop?}
%gives an overview of the interview study contacted as part of the \emph{Precog} study: 

%%%%%%%%%%%%%%%%%%%%%%%%%%%%%%%%%%%%%%%%%%%%%%%%%%%%%%%%%%%%%%%%%%%%%%%%

\subsection{Preparation of interviews}
\begin{figure}[!htbp]
  \centering
 \includegraphics[width=\linewidth]{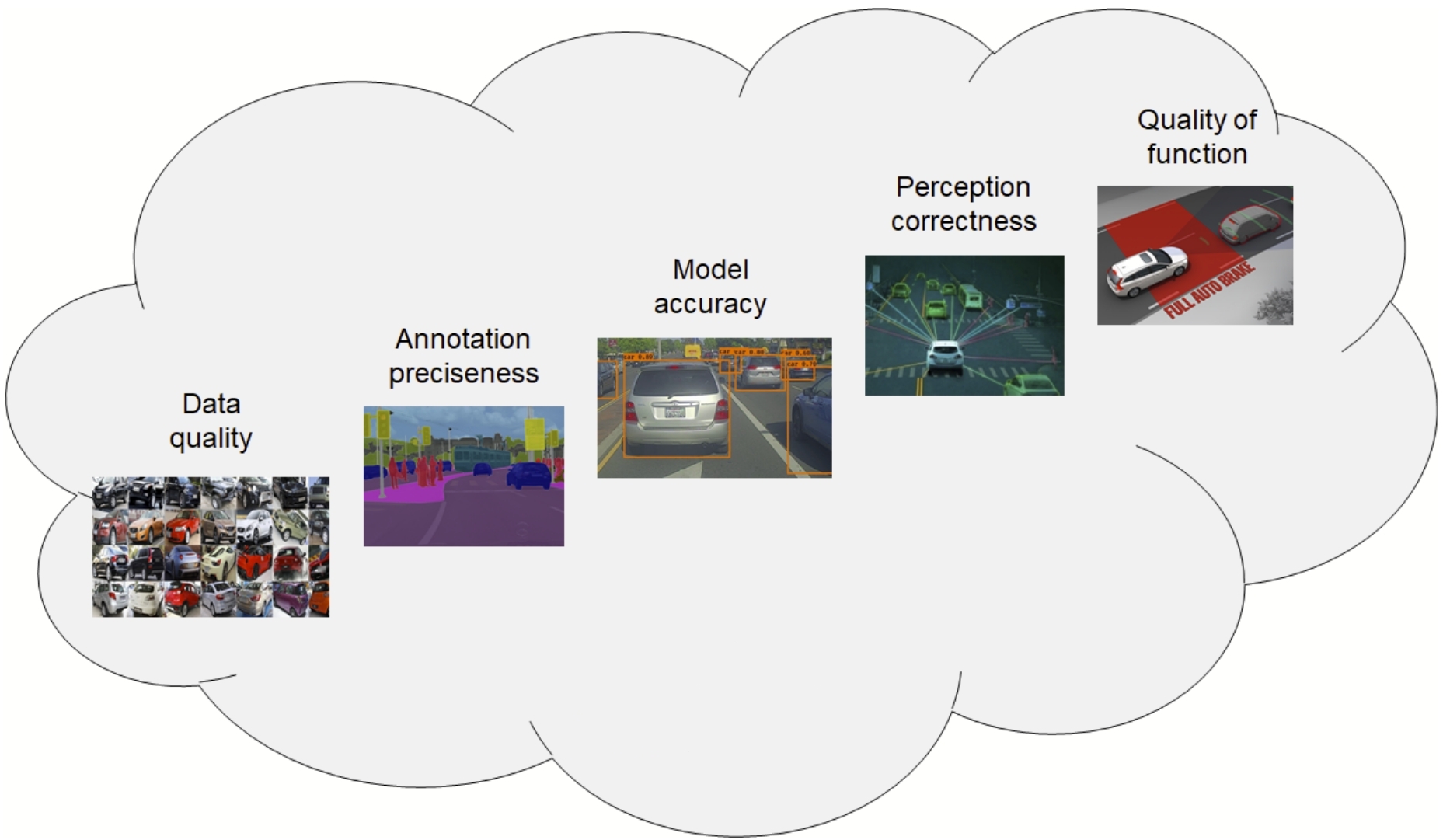}
  \caption{Building a safety case for automotive perception software}
  \label{fig:interview_2}
  \vspace{-1em}
\end{figure}

The guiding theme of the interviews was the exploration of requirements and software engineering practises for automotive perception software that incorporates ML. 
Based on a-priori formulated research questions, the involved researchers created an interview guide\footnote{accessible in the replication package \url{https://doi.org/10.7910/DVN/HCMVL1}} with nine parts, whereof two parts were optional depending on available time.
The first part aimed at collecting information about the background of the interviewees. 
In the second part, we used a diagram$^4$ of a typical architecture for automotive perception systems and asked the interviewees to position themselves in terms of the architecture. 
The aim was to find common ground with the interviewees and to understand their typical field of work.
In the next part, we concentrated on discussing the processes used to ensure correctness of the perception system. 
Figure~\ref{fig:interview_2} was used to discuss with the interviewees how requirements on the quality of the function relate to requirements on data and annotation needs. 
%\hl{We used safety as an example of a quality aspects for ML models here because safety is a key property of automotive perception systems.} 
%We were especially interested in learning how safety and quality requirements on the perception software components influence requirements on data used for training machine learning components. 
Part four of the interview guide contains questions regarding the safety case, and how ML models can become a part of the safety case for software components. 
The next part contained questions regarding the context description in which a safety case of the software component is valid. 
Part six contained questions about the ecosystem and business environment in which the perception software is developed. Here, we were curious in how safety critical software that includes ML models are developed together with partners and suppliers and how expectations in the software, the ML model and data are communicated. 
The next two parts were optional and contained questions regarding quality trade-offs. 
The final part closed the interview and allowed the interviewees to suggest improvements and additional interviewees.
\paragraph*{Sampling strategy}
Our sampling strategy was a mix of purposeful and snowball sampling. For the latter, we sent open calls to contacts in the Swedish automotive industry, including the partners involved in the \emph{Precog} study, and explicitly asked both before and after the interview for additional contacts we could interview.
With respect to purposeful sampling we aimed to interview practitioners from industry who had experience with developing software for automotive perception systems, data science, machine learning, requirement engineering, and safety engineering. 
Because we knew that we cannot find all these qualifications in a single person, we decided to conduct group interviews to cover all desired competencies. 
We also hoped to enable better discussions in a small group settings. 

%\begin{figure*}[!htbp]
%\centering
%\begin{subfigure}{.5\textwidth}
%  \centering
% \includegraphics[width=\linewidth]{figs/interview_1.jpg}
%  \caption{Positioning along an architecture diagram}
%  \label{fig:interview_1}
%\end{subfigure}%
%\begin{subfigure}{.5\textwidth}
%  \centering
%  \includegraphics[width=0.9\linewidth]{figs/interview_2.jpg}
%  \caption{Process and correctness for data and annotations}
%  \label{fig:interview_2}
%\end{subfigure}
%\caption{Figures used during the interviews}
%\label{fig:interview_figures}
%\end{figure*}

%%%%%%%%%%%%%%%%%%%%%%%%%%%%%%%%%%%%%%%%%%%%%%%%%%%%%%%%%%%%%%%%%%%%%%%%

\subsection{Data collection}
We conducted two campaigns for data collections: Group interviews and a validation workshop.
\subsubsection*{Interviews}
In seven interview session we interviewed 19 participants from five companies. Three international automotive OEMs, one supplier, and one annotation company participated in the study. In each session, at least two scientists were present. The interview sessions were conducted between December 2021 and April 2022 via Microsoft Teams and took 90 to 120 minutes. A list of the participants for each interview is given in Table~\ref{tab:participants}. All interviews were recorded, automatically transcribed, anonymised, mistakes in the automatic transcripts manually corrected, and the final transcripts stored as spreadsheets for further analysis.

In the invitation e-mail to each group of interviewees we informed about the research project's goal, financial support, and duration. In the beginning of the interview, we provided some formal information such as data privacy compliance and permission for recording and data processing. Then, the interview guide was followed. For each section of the interview guide one interviewer asked the questions, while the remaining interviewers observed and took notes.\par 
The interviews were semi-structured using the set of predetermined open-ended questions formulated in the interview guide. However, we allowed deviations from the questions and the order of questions to facilitate discussions, also among the group of interviewees. 

\begin{table}[t]
\centering
\footnotesize
\caption{Overview of conducted interviews}
\label{tab:participants}
\begin{tabular}{cll}
\hline
\multicolumn{1}{c}{\textbf{\begin{tabular}[c]{@{}c@{}}Inter-\\ view\end{tabular}}} & \multicolumn{1}{l}{\textbf{Field of work}} & \multicolumn{1}{l}{\textbf{Participants}} \\ \hline
\rowcolor[HTML]{EFEFEF} 
A & Object detection & Product owner \\
B & Autonomous Driving & \begin{tabular}[l]{@{}l@{}}Product owner \\ Test engineer \\ AI engineer \\ Software developer \end{tabular} \\
\rowcolor[HTML]{EFEFEF} 
C & Vision systems & \begin{tabular}[l]{@{}l@{}}System architect \\ Product owner \\ Requirement engineer \\ Deep learning engineer \end{tabular} \\ 
D & AD and ADAS & \begin{tabular}[l]{@{}l@{}}System engineer \\ Manager AD$^1$ \end{tabular} \\
\rowcolor[HTML]{EFEFEF} 
E & Testing and validation AD$^1$ &\begin{tabular}[l]{@{}l@{}}System architect \\ Product owner \\ Product owner \\ Compliance officer \\ Data Scientist \end{tabular} \\
F & Data annotations &\begin{tabular}[l]{@{}l@{}} AI engineer \\Data scientist \end{tabular} \\
\rowcolor[HTML]{EFEFEF} 
G & Autonomous Driving & \begin{tabular}[l]{@{}l@{}} System safety engineer \end{tabular} \\ \hline
\multicolumn{3}{l}{\scriptsize The participants' IDs are random and not shown to ensure confidentiality.} \\
\end{tabular}
\vspace{-1em}
\end{table}

%%%%%%%%%%%%%%%%%%%%%%%%%%%%%%%%%%%%%%%%%%%%%%%%%%%%%%%%%%%%%%%%%%%%%%%%

\subsection{Data analysis}
We applied a mixed coding strategy for analysing the qualitative data obtained through the interviews. Mixed coding strategies can be suitable for settings in which coding of in-depth interviews is conducted by teams using a shared platform, such as in our case Microsoft Office 365 \cite{Deterding2021}. Each coding team consisted of at least three researchers who conducted the first round of coding together. The team started with a number of high-level deductive codes which were based on the interview questions and researchers' experience. Then, while applying the deductive codes, new codes emerged as part of an inductive coding scheme. 
These emerging codes were added to a shared list used by all coding teams. 
After five interviews, we observed saturation by noticing that not many new inductive codes emerged in the following interviews. In a second round of coding, a new group of at least two researchers first revisited each interview and then applied the final list of emerged codes. Afterwards, pattern coding was used to identify emerging themes and sub-categories \cite{Saldana2013}. Finally, each statement of the interviewees was assigned to the identified sub-categories. The final codes of each interview were reviewed by an additional independent researcher.

%%%%%%%%%%%%%%%%%%%%%%%%%%%%%%%%%%%%%%%%%%%%%%%%%%%%%%%%%%%%%%%%%%%%%%%%

\subsection{Result validation}
%A first step towards result validation was a peer review of the emerged codes and identified themes by researchers who were not involved in the data analysis.
Towards the end of the \emph{Precog} study in April 2022, we conducted a 2.5-hour workshop with 25 participants from industry, where five participants were also interviewees. The aim of the workshop was to validate and discuss the preliminary findings of the interview study. The workshop was conducted on-site in Göteborg, Sweden, but with the possibility of connecting remotely using the Zoom conference software.
%The workshop was held in a hybrid setting with 25 participants joining either on-site or through the Zoom conferencing software. 
During the workshop we first presented the identified sub-categories for each theme on the online whiteboard platform Miro. One theme at a time we let the participants vote on which categories are most important or relevant to them. The participants could give a maximum of one vote to each category and they were allowed to vote for more than one category in each theme.
%We allowed each participant to vote several times for each themes, but only once per sub-category.  
We use these results to gauge the relative importance of our discovered sub-themes according to participating practitioners.

%%%%%%%%%%%%%%%%%%%%%%%%%%%%%%%%%%%%%%%%%%%%%%%%%%%%%%%%%%%%%%%%%%%%%%%%
%%%%%%%%%%%%%%%%%%%%%%%%%%%%%%%%%%%%%%%%%%%%%%%%%%%%%%%%%%%%%%%%%%%%%%%%
%%%%%%%%%%%%%%%%%%%%%%%%%%%%%%%%%%%%%%%%%%%%%%%%%%%%%%%%%%%%%%%%%%%%%%%%

\section{Results} \label{sec:results}
This section presents the identified themes around challenges when specifying data and annotations of data for automotive perception software that incorporates ML. Furthermore, this section presents themes regarding the ability of automotive industry's ecosystem and business models to handle data-intensive developments, such as the design, developing and deployment of software that incorporates ML models. 

%%%%%%%%%%%%%%%%%%%%%%%%%%%%%%%%%%%%%%%%%%%%%%%%%%%%%%%%%%%%%%%%%%%%%%%%

\subsection{RQ1: The ability to specify data for the development of automotive perception software}

Unlike conventional software systems in which rules specify the desired behaviour, a software component that incorporates ML infers these rules from data. Therefore, data plays a prominent role when trying to ensure correct behaviour of ML-based software for perception systems. If the data provided to an ML algorithm is biased, the resulting ML model will learn that bias, and consequently the decisions of the perception software will be biased.

During the interviews we wanted to learn about the interviewees' understanding of ``data quality". We furthermore asked about the processes and ways of working used in relation to specifying and collecting data for the development of data-intensive software. 

Figure~\ref{fig:cause_effect_all} illustrates the three major themes that we identified in the interviews regarding the ability to specify data used for the development of software for perception systems: 1) Data collection, 2) Processes \& Way of Working, and 3) Data quality. For each themes, a set of sub-categories was identified. Table~\ref{tab:RQ1} lists these categories. The indicated score is the percentage of votes given for a sub-category out of all votes given for a theme.

%\eric{Integrating the table into the fishbone will safe space and increase clarity (more than removing the figures).}\eric{how did you decide which subthemes to cover? Based on Score, I hope?}
\begin{table}[!htbp]
\centering
\footnotesize
\caption{Overview of all themes and sub-categories for RQ1: Challenges affecting the ability to specify data used for data-intensive software development (n=46), n is the number of submitted votes}
\label{tab:RQ1}
\begin{tabular}{lllc}
\hline
\multicolumn{2}{c}{\textbf{ID}}                              & \multicolumn{1}{c}{\textbf{Description}}  & \multicolumn{1}{c}{\textbf{Score}}                                                                           \\ \hline
\rowcolor[HTML]{EFEFEF} 
{\textbf{D1}} &  \multicolumn{3}{l}{\cellcolor[HTML]{EFEFEF}\textbf{Data collection}}                  \\
\rowcolor[HTML]{EFEFEF} 
                                   & -I                                                    & data selection & 13\%                                                \\
\rowcolor[HTML]{EFEFEF} 
                                   & -II                                                     & simulation & 9\%                                     \\

\rowcolor[HTML]{EFEFEF} 
                                   & -III                                                     & data collection & 7\%                                                  \\
\rowcolor[HTML]{EFEFEF} 
                                   & -IV                                                    & metadata & 4\%                                                      \\
\rowcolor[HTML]{EFEFEF} 
                                   & -V                                                    & experimentation & 2\%                                                      \\
\rowcolor[HTML]{EFEFEF} 
                                   & -VI                                                    & synthetic data & 2\%                                                      \\
\textbf{D2}                                             & \multicolumn{3}{l}{\textbf{Processes and Ways of Working}}                                                                    \\
                                   & -I                                                      & data specification & 7\%                                                        \\
                                   & -II                                                     & data requirements & 4\%                                             \\
                                   & -III                                                     & data verification & 4\%                                                \\
                                   & -IV                                                     & data cleaning & 2\%                                                \\
                                   & -V                                                     & data collaboration & 2\%                                                \\
                                   & -VI                                                     & data storage & 0\%                                                \\
\rowcolor[HTML]{EFEFEF} 
\textbf{D3}                                  & \multicolumn{3}{l}{\cellcolor[HTML]{EFEFEF}\textbf{Data quality}}                                             \\
\rowcolor[HTML]{EFEFEF} 
                                   & -I                                                      & data variation & 11\%                                                                 \\
\rowcolor[HTML]{EFEFEF} 
                                   & -II                                                    & bias & 9\%                                                                   \\
\rowcolor[HTML]{EFEFEF} 
                                   & -III                                                    & future-proof dataset & 9\%                                                          \\
\rowcolor[HTML]{EFEFEF}                                    
                                   & -IV                                                    & data quality & 7\%                                                          \\
\rowcolor[HTML]{EFEFEF}                                    
                                   & -V                                                   & data correctness & 4\%                                                          \\ 
\rowcolor[HTML]{EFEFEF}    
                                   & -VI                                                   & data re-usability & 4\%
                                   \\ 
\rowcolor[HTML]{EFEFEF}    
                                   & -VII                                                   & data maintainability & 0\%
                                   \\                                    
 \hline
%\multicolumn{4}{l}{\begin{scriptsize}\begin{tabular}[l]{@{}l@{}}$^1$The score (s) has been calculated as the votes (v) \\ given for a sub-theme (i) normalised over the total \\ numbers of votes for this theme ($\mathrm{n_j}$) \\ multiplied by 100, i.e., $s_i=100\cdot v_i/\mathrm{n_j}$. \end{tabular}\end{scriptsize}} 
\end{tabular}
%\vspace{-1em}
\end{table}

%%%%%%%%%%%%%%%%%%%%%%%%%%%%%%%%%%%%%%%%%%%%%%%%%%%%%%%%%%%%%%%%%%%%%%%%

\subsubsection{Data collection}
Ensuring the correct behaviour of software that contains ML models requires a highly data-driven development. The participants of the validation workshop ranked the sub-category \emph{data selection} as a key aspect of data collection to ensure safe and correct behaviour. Often, the right dataset as input to ML algorithms is found through a set of iterations between the ML experts and the data scientists. For example, uncertainty measures can be used to decide which additional data needs to be selected to reduce uncertainty:
\interviewquote{To ensure some form of safety measures from the model, we produce \textbf{uncertainty estimations from outputs}. Those are used in the data selection of course to look for what type of data are we uncertain. What do we need to learn more from?}{Interviewee~B-I}
\noindent Because finding the right data can be expensive, development teams try to use simulations for training and validation purposes. The participants even consider data from simulation more important than data originating from planned experiments, such as test drives.
\interviewquote{Furthermore, \textbf{simulations are very often an integral part of test strategies} for machine learning based systems.}{Interviewee~C-IV}
\noindent There can be two reasons for preferring data from simulation: It is often significantly cheaper to obtain data through simulations. And, it is possible to obtain data for rare case scenarios that could be impossible to obtain in real world experiments:
\interviewquote{Especially like for the \textbf{rare case scenario that is not really easy} to replicate in real world, so we cannot.}{Interviewee~E-II}
% 

%%%%%%%%%%%%%%%%%%%%%%%%%%%%%%%%%%%%%%%%%%%%%%%%%%%%%%%%%%%%%%%%%%%%%%%%

\subsubsection{Processes and Way of working}
We asked during the interviews which processes and ways of working in regard to data used in the development of safety-critical data-intensive software is being used. Most safety standards, such as ISO 26262, rely on the correctness of processes to build up a safety case for a product. Therefore, it is important to define processes that ensure the safety of software with ML components.

The most important capability, defined formally through a process or informally through a way of working, is the creation of data specifications. We saw earlier that data selection is the most important activity in a data driven development process. Data specifications are a logical prerequisite to data selection. Ideally, data specifications precise the requirements set on the data in relation to for example physical data properties, data quality, and quantitative targets \cite{Vogelsang2019}. But it is often quite unclear what a data specification should entail:
\interviewquote{It's \textbf{very different how you write a data specification} [...] it's hard to know what the future expects and what type of classes we want and how we do want to combine certain objects.}{Interviewee~B-I}
\noindent An iterative processes can be used to find the final data specification of the system: 
\interviewquote{It's more of a sort of a \textbf{data driven and then statistical analysis of the data in a continuous way}. So we start with logging data, annotating it, training or models, and so on. And then we can also draw some statistics. OK, how is the class balance in this dataset? How does that affect the per class accuracy? Do we need to look for more? And then we can of course feedback that to the data selection team and they can start looking for certain classes, for example.}{Interviewee~B-I}
\noindent The statement shows that a specification typically consists of a set of requirements such as accuracy, balance, etc. We found however that the exact scope of a data requirement is not entirely clear. Data requirements can for example describe desired probability distributions and quantity of the data:
\interviewquote{We write documents, word documents, basically where \textbf{we describe the distribution of the data and the quantity of the data} that needs to be collected.}{Interviewee~C-II}
\noindent Data requirements can also entail specific data quality aspects, such as pixel density, brightness, size of bounding boxes, etc. In both understandings of data requirements, they allow for data verification. Data verification means checking that the data is representative for the desired "real-world" scenario.

Data specifications and data requirements entailed within the specifications are key enablers for many companies to collaborate in data collection and processing for example with supplier companies: 
\interviewquote{We have a 3rd party company \textbf{driving around all this mileage} and collecting data. They want you to send over that data to them for doing the simulations. \textbf{And then they will put their requirements on what sort of data we are collecting.}}{Interviewee~D-I}
\noindent A final data specification that describes the utilised data can be a key input towards a safety case of a ML-based perception system and allows for verification of design decision. 
%\eric{Perhaps verification should have been mentioned?}
%%%%%%%%%%%%%%%%%%%%%%%%%%%%%%%%%%%%%%%%%%%%%%%%%%%%%%%%%%%%%%%%%%%%%%%%

\subsubsection{Data quality}
Data requirements often entail some desired data quality aspect. Interestingly, the most important data quality aspects mentioned by the interviewees do not describe physical properties of data, such as pixel density, contrast, resolution, brightness, etc., but instead focus on the represented information in the data.

Because data selection was identified as the most important aspect for data collection, it is not surprising that data variation has been chosen by the study participants as the most important data quality characteristic, even before data correctness. Data variation is directly causally related to bias in data; a lack of data variation will result in bias, which can propagate into the ML model. 

A challenge regarding data variation is the definition of KPIs, or in general measures of variety. How do you measure variety, and when do you know that your data has enough variety?
\interviewquote{How would you divide that space and define it in a way that allows a measure of have I covered not only enough children, but also \textbf{enough variety} of children [as vulnerable road users]?}{Interviewee~F-I}
\noindent Both data variation and data correctness require however an a-priori understanding of the environment in which the software will be deployed. If the operational domain is unknown, it will be difficult to describe what variety entails:
\interviewquote{You need to \textbf{understand the distribution of where to collect data and that requires an understanding of where the function in the end will be used.}}{Interviewee~C-I}
\noindent Collecting and processing data is often a costly part of the development process. Therefore, the re-usability, and future-proofness of data are considered important data quality aspects. 
 
\interviewquote{What do we need to ensure to make use of the data we've collected up to now? I mean, \textbf{how do we make sure we don't have to start from scratch?}}{Interviewee~E-I}
	
%%%%%%%%%%%%%%%%%%%%%%%%%%%%%%%%%%%%%%%%%%%%%%%%%%%%%%%%%%%%%%%%%%%%%%%%

\subsection{RQ2: The ability to specify annotations for data used in automotive perception software}

The development of ML models often relies on supervised learning which requires annotated datasets. There are approaches towards automatised annotation of data\footnote{automated annotation for example has been attempted for images \cite{Adnan2021}, for textual data \cite{Ding2018}, or videos \cite{Berg2019}} but these approaches regularly do not succeed in replacing human-in-the-loop annotators \cite{Wu2022}. Because annotation plays a major role in the development of perception software that incorporates ML, both from a performance and cost point-of-view, we investigated which challenges practitioners encountered when specifying annotations for data.

Figure~\ref{fig:cause_effect_all}  shows the three major themes we identified within annotation challenges. Refined sub-categories for each theme together with the score they received in the validation workshop are listed in Table~\ref{tab:RQ2}.
%\begin{figure}[!htbp]
%    \centering
%    \includegraphics[width=\linewidth]{figs/cause_effect_annotation.pdf}
%    \caption{Cause-Effect diagram of themes regarding the ability to specify data annotations for data-intensive software development.}
%    \label{fig:cause_effect_annotation}
%\end{figure}
%
\begin{table}[!htbp]
\centering
\footnotesize
\caption{Overview of all themes and sub-categories for RQ2: Challenges affecting the ability to specify annotations for data used for data-intensive software development (n=31)}
\label{tab:RQ2}
\begin{tabular}{lllc}
\hline
\multicolumn{2}{c}{\textbf{ID}}                              & \multicolumn{1}{c}{\textbf{Description}}  & \multicolumn{1}{c}{\textbf{Score}}                                                                           \\ \hline
\rowcolor[HTML]{EFEFEF} 
{\textbf{A1}} &  \multicolumn{3}{l}{\cellcolor[HTML]{EFEFEF}\textbf{Annotation costs}}                  \\
\rowcolor[HTML]{EFEFEF} 
                                   & -I                                                    & annotation cost & 3\%        \\
\textbf{A2}                                             & \multicolumn{3}{l}{\textbf{Annotation quality}}                                                                    \\
                                   & -I                                                      & annotation consistency & 16\%                                                        \\
                                   & -II                                                     & annotation correctness  & 13\%                                             \\
                                   & -III                                                     & annotation quality  & 13\%                                                \\
                                   & -IV                                                     & annotation validation & 10\%                                                \\
                                   & -V                                                     & annotation re-usability & 6\%                                                \\
                                   & -VI                                                     & annotation precision & 3\%                                                \\
                                   & -VII                                                     & pixel precision & 0\%                                                \\                                   
\rowcolor[HTML]{EFEFEF} 
\textbf{A3}                                  & \multicolumn{3}{l}{\cellcolor[HTML]{EFEFEF}\textbf{Guidelines \& Specification}}                                             \\
\rowcolor[HTML]{EFEFEF} 
                                   & -I                                                      & ground truth & 13\%                                                                 \\
\rowcolor[HTML]{EFEFEF} 
                                   & -II                                                    & annotation specification & 10\%                                                                   \\
\rowcolor[HTML]{EFEFEF} 
                                   & -III                                                    & annotation guidelines & 6\%                                                          \\
\rowcolor[HTML]{EFEFEF}                                    
                                   & -IV                                                    & labelling & 6\%                                                          \\
                                    
 \hline
\end{tabular}
%\vspace{-1em}
\end{table}

%%%%%%%%%%%%%%%%%%%%%%%%%%%%%%%%%%%%%%%%%%%%%%%%%%%%%%%%%%%%%%%%%%%%%%%%

\subsubsection{Annotation costs}
The final cost of annotation often is described as a trade-off between annotation quality and quantity. All interview partners agree that the cost of annotation rises exponentially with the level of annotation quality and linearly with the quantity of annotated data. Both a higher quality of annotations and a higher quantity of annotated data can result in performance increases of the trained ML model:
\interviewquote{And there's some kind of scaling. So if you don't have a lot of frames [annotated], your model performance will be way worse. And if you would have a higher quality then you would of course get a higher performance. But at some point there's a diminishing of returns.}{Interviewee~F-II}
\noindent Given a fixed budget for annotation, a trade-off is described as choosing between either high quantity of data with low quality annotations, or vice versa: 
\interviewquote{[...] OK if you have higher quality then maybe you can do with less [annotated] data instead. And then you save money from one perspective. But yeah, it takes longer time to do one precise annotation. So that \textbf{you have to balance} the two a bit.}{Interviewee~C-II}

%%%%%%%%%%%%%%%%%%%%%%%%%%%%%%%%%%%%%%%%%%%%%%%%%%%%%%%%%%%%%%%%%%%%%%%%

\subsubsection{Annotation quality}
Annotations are regarded a \emph{quality aspect of data}, but there is no clear distinction in what this quality aspect entails. It can refer to qualitative aspects of annotations such as the precision of annotation boxes or the correctness of annotations. But it can also refer to quantitative aspects, such as the amount of annotated data or how many features are labelled within a single frame. Furthermore, there is no distinct definition of \emph{annotation quality}. The uncertainty in the definition of annotation quality can stem from the inability to define clear quality metrics which can be used for setting requirements on the annotation (similar to the inability to define metrics for data variety as a data quality criteria):
\interviewquote{And just to add, it's not really clear \textbf{how we can measure the quality of annotations itself} as well. Like how to make sure that even like if you put a requirement on the annotations and they have reached their level of quality that you asked for.}{Interviewee~D-I} 
\noindent The uncertainty in the specifications of annotation can results in uncertainties and even consistency problems in the annotated dataset: 
\clearpage%
\interviewquote{I think it's a pitfall. Maybe that it's easy to look at, you know, like pixel precision: ah, you are two pixels out of the actual border of the object here, but I think maybe we've seen a bigger problem that \textbf{one type of object has one class in one label and another class in the next image} because hundreds of annotation people have interpreted the specifications differently. I think for us that's a bigger problem in annotation quality than the pixel precision.}{Interviewee~B-II}
%
%\interviewquote{If we start with precision versus consistency: [...] It's much easier to say that, OK, we need more precise pixel precision for each class. [...] Consistency is one thing that needs to be quality checked consistently [...] to make sure that OK we are labelling certain objects in all of the frames, not the scenarios, no matter the domain and so on.} {Interview~B}
\noindent Being able to provide consistent annotation has been ranked as the most pressing challenge by the validation workshop participants. There are approaches described in literature that in theory allow for testing consistency between annotators (for example \cite{Wang2019}), but it seems not to be used in practise yet. At the same time, pixel precision is not considered an equally pressing challenge:
\interviewquote{It's \textbf{much easier to solve the preciseness problem than the consistency problem}. It's our experience.}{Interviewee~B-I}
\noindent Furthermore, as a consequence of high annotation costs there is a desire to re-use annotated data:
\interviewquote{[...] we still would like to use that data because I \textbf{paid a lot of money to annotate it} and then we can do different things [...] that actually contributes to robustness. }{Interviewee~A-I}

%%%%%%%%%%%%%%%%%%%%%%%%%%%%%%%%%%%%%%%%%%%%%%%%%%%%%%%%%%%%%%%%%%%%%%%%

\subsubsection{Guidelines \& Specification}
In theory, specifications serve two purposes: A \emph{requirements specification} documents the requirements that need to be fulfilled by an item and a \emph{technical specification} can document the features of an item that fulfil the desired requirements. They are key communication artefacts between OEMs and suppliers or other external companies \cite{Shishodia2019}. More and more, annotations are conducted by external companies. Yet, for the process of annotating data, clear requirements specifications often are not formulated. This results in ambiguous expectations on the resulting annotations:
\interviewquote{[A]s soon as something is not explicitly stated, it's they [the annotators] don't know what to do because we experience [that] they can extrapolate what they know, but \textbf{there's still a lot of these either ambiguous or hard to tell scenarios that are usually quite unspecific}.}{Interviewee~F-I}
\noindent As a consequence of missing requirements specification for annotation, there is ambiguity in the guidelines that describe the annotation process. For example, how should uncertainties during annotation be handled? 
\interviewquote{But which are still sort of not properly described or it's not properly defined how to act in that situation. [...] What is annotated should it be [marked as] correct, or is it better to mark [it] as unclear? \textbf{Maybe it's worse to be wrong than to be not sure in this case.}}{Interviewee~D-I}
\noindent The question about certainty in the annotation is relevant, because many annotation processes apply time constraints towards the annotators. With a fixed time budget, for example a maximum of 20 seconds per image, the quality of the annotations might be significant worse than if no time constraint is applied in the annotation process because the annotator has more time to concentrate on the details of each frame:
\interviewquote{In one way \textbf{it's a trade off of the time taken to peruse annotations versus the actual quality}. So I mean, for example, if we say you have one minute to do each task, or if you have, how however much time you need. We would naturally assume that the later approach will enable us to create more detailed annotations.}{Interviewee~F-I}
\noindent Finally, without \emph{annotation specifications}, and therewithin information about the annotation process, the company receiving the annotated data from a subcontractor cannot judge on the reliability of the annotated data:
\interviewquote{I mean it depends on what, what if they have it in house, or if they outsource it and so on. But when it comes down to the labelling, that means \textbf{they have to tell us what their actions or processes [are] and then how they guarantee the quality or efficiency of the labelling}. And then we have to take that judgement.}{Interviewee~D-I}
\noindent The ability to provide annotation specifications is also of significance towards safety evidence. Because the annotation process has influence on the final performance and correct behaviour of a ML model that is part of a perception software, it needs to be clearly documented as part of a safety case.
%%%%%%%%%%%%%%%%%%%%%%%%%%%%%%%%%%%%%%%%%%%%%%%%%%%%%%%%%%%%%%%%%%%%%%%%

%\subsection*{Unused quotes}

%\interviewquote{So this is only on the data collected and then we have another process that we call data curation. This is also [...] where he and his team select which data out of this pretty huge number of hours we have collected [...] we actually annotate?}{Interview~C}

%\interviewquote{Basically that's not the completely clear answer to that. And also about the the first box, data quality. It could be both. I interpret it as both the quality of the data itself like that, it's so for instance is not broken and it's logged perfectly and so on. But also like that we select some of the data from the the whole data that we log because basically it's not rational to send all the collected data for annotations.}{Interview~E}

%\interviewquote{I have a very similar view also coming from other requirements work in other parts of the industry. Like usually you start with the customer view or customer value and the customer doesn't really care if we have good perception, they care about what the function does, in the end, and that's why we end up needing correct perception. And to get the correct perception we need model accuracy and to get to even understand if we have model accuracy we need good annotation and good data.}{Interview~C}

%%%%%%%%%%%%%%%%%%%%%%%%%%%%%%%%%%%%%%%%%%%%%%%%%%%%%%%%%%%%%%%%%%%%%%%%

\subsection{RQ3: Automotive industry's ecosystems and business models for data-intensive software developments}
The previous section highlighted the importance of data and annotation specifications for the development of software that incorporates ML components in the automotive industry. Requirement documents, for example requirements specifications, play a major role in steering the process flow between OEMs and their suppliers \cite{Allmann2006}. Typically, major parts of a vehicle's software are developed externally through suppliers in agreement with the requirement specifications they receive from the OEMs. However the typical specification driven approach fails in scenarios of complex system development, for example components with significant linkage between different systems \cite{Bach2017}, and highly data driven developments such as deep learning \cite{Kaiser2021}.

In this section we investigate causes of the challenges that affect the ability of the automotive industry's ecosystems and business models to handle data intensive development. Based on the conducted interviews and the validation workshop we identified three major themes as shown in Figure~\ref{fig:cause_effect_all} and twelve sub-categories listed in Table~\ref{tab:RQ3}.
\begin{table}[!htbp]
\centering
\footnotesize
\caption{Overview of all themes and sub-categories for RQ3: Challenges affecting the ability of the automotive industry's ecosystems and business models to handle data-intensive software development (n=46)}
\label{tab:RQ3}
\begin{tabular}{lllc}
\hline
\multicolumn{2}{c}{\textbf{ID}}                              & \multicolumn{1}{c}{\textbf{Description}}  & \multicolumn{1}{c}{\textbf{Score}}                                                                           \\ \hline
\rowcolor[HTML]{EFEFEF} 
{\textbf{B1}} &  \multicolumn{3}{l}{\cellcolor[HTML]{EFEFEF}\textbf{Business environment}}                  \\
\rowcolor[HTML]{EFEFEF} 
                                   & -I                                                    & value chain & 7\%                                                \\
\rowcolor[HTML]{EFEFEF} 
                                   & -II                                                     & ecosystem & 4\%                                     \\

\rowcolor[HTML]{EFEFEF} 
                                   & -III                                                     & feedback & 4\%                                                  \\

\textbf{B2}                                             & \multicolumn{3}{l}{\textbf{Contracts \& Infrastructure}}                                                                    \\
                                   & -I                                                      & tools & 22\%                                                        \\
                                   & -II                                                     & contracts & 7\%                                             \\
                                   & -III                                                     & negotiations & 4\%                                                \\

\rowcolor[HTML]{EFEFEF} 
\textbf{B3}                                  & \multicolumn{3}{l}{\cellcolor[HTML]{EFEFEF}\textbf{Shared responsibility}}                                             \\
\rowcolor[HTML]{EFEFEF} 
                                   & -I                                                      & transparency & 20\%                                                                 \\
\rowcolor[HTML]{EFEFEF} 
                                   & -II                                                    & collaboration & 15\%                                                                   \\
\rowcolor[HTML]{EFEFEF} 
                                   & -III                                                    & legal aspects & 7\%                                                          \\
\rowcolor[HTML]{EFEFEF}                                    
                                   & -IV                                                    & risk of litigation & 4\%                                                          \\
\rowcolor[HTML]{EFEFEF}                                    
                                   & -V                                                   & business model & 4\%                                                          \\ 
\rowcolor[HTML]{EFEFEF}    
                                   & -VI                                                   & crowd sourcing & 2\%
                                   \\ 
                                
 \hline
\end{tabular}
%\vspace{-1em}
\end{table}
%
%%%%%%%%%%%%%%%%%%%%%%%%%%%%%%%%%%%%%%%%%%%%%%%%%%%%%%%%%%%%%%%%%%%%%%%%

\subsubsection{Business environment}

The ``conventional" value chain in the automotive industry is based on sourcing suppliers which provide the OEMs with the technology needed for their products. Because of the emergence of data-intensive software in vehicles and the agile transformation that embraced the automotive industry\footnote{where the agile transformation can be an effect of the introduction of data-intensive software \cite{Hoda2018}}, this type of partnership in the value chain cannot work anymore:
\interviewquote{Yeah, \textbf{partnerships is a very hot word instead of suppliers}. So and I think that captures it, it cannot be a classical, what's it called, purchaser supplier relation, where you just write what you want and then you get it. This is like the sourcing new managers dream world. \textbf{They get the blueprint, the drawing and then they go to five different vendors and then they pick the cheapest one. But it doesn't work like that at all}. It was hard already with classical software. It's like impossible here, so it needs to be more of a partnership where it's possible to have a dialogue and to iterate without, you know, having to go through the whole commercial process once more, so it needs to be a bit more loose ended.}{Interviewee~B-II}
\noindent Instead of traditional sourcing, OEMs seek project partnerships for example with major technology companies. Alternatively, they partner up (and eventually buy) technology start-up companies, especially around data collection, annotation services, or automatic driving in general. The reason is that OEMs aim at integrating suppliers closer into their own development:
\interviewquote{[...] we did have a pretty extensive thorough sourcing project when we choose the suppliers that we're working with now. And that was a major factor. We didn't just pick the cheapest one or the one that we thought had their absolute best, you know, maybe accuracy. \textbf{We picked the one that we feel that we can work with these people and you know, we can have a dialogue and it's possible to make adaptions without getting into a commercial discussion}. You know, from the first minute. It's more like a partnership spirit or setup.}{Interviewee~B-II}
\noindent The comment that an OEM is willing to pay a premium for joint teams development stems from the need of regular feedback in complex system development. Traditionally, suppliers only allow limited access to the development details in \emph{joint reviews}. However, these discrete feedback points are not sufficient for an agile mindset required for data-intensive software projects: \vspace{10em}
\interviewquote{Normally with the suppliers, if you assume that they are a normal supplier, you cannot see those stuff unless you go to some kind of on site joint reviews because \textbf{it's IP basically}.}{Interviewee~G-I}
\noindent We identified two major causes for the need of continuous feedback loops: First, the operational design domain (ODD), i.e., the context in which the system can operate as expected, is initially not entirely known. Instead, the true ODD is jointly ``discovered" during the development. 
\interviewquote{[...] it's basically giving some feedback to the system design, function design and so on to modify the function and introduce the limitations to the function which we name it ODD. \textbf{An iterative loop is going on and on and on until all of these triggering events are acceptable basically}.}{Interviewee~G-I}
\noindent Second, the desired quality level of the software, especially in relation to safety and security, can only be achieved by continuously improving for example the data selection or annotations, even after the software has been deployed.

%\interviewquote{Another thing we're doing [...] is that we're actually looking together with manufacturers what is the worst case you would you expect to see in the field so that we [...] span the whole range of images you can expect.}{Interview~E}

%%%%%%%%%%%%%%%%%%%%%%%%%%%%%%%%%%%%%%%%%%%%%%%%%%%%%%%%%%%%%%%%%%%%%%%%

\subsubsection{Contracts \& Infrastructure}
New forms of cooperation require new ways of \emph{negotiating with} and \emph{contracting of} suppliers, as well as tools that support continuous collaborations. A typical contract can contain function requirements that a supplier needs to fulfil. However, for data driven and probabilistic systems, function requirements are often very abstract and provide little guidance for system development: %
\interviewquote{What I get [from customers] are usually those very abstract function requirements that I should never miss a single pedestrian, ever.[...] So what I try to do is that I break down and do decomposition [until] I \textbf{end up with feasible requirements on each component}.}{Interviewee~C-III}
\noindent The key issue with contracting of suppliers for data driven development is the definition of success. It seems difficult to define clear development goals due to the iterative nature of data driven development:
\interviewquote{Like \textbf{what's the definition of success} when you're really building a [machine learning] model and going back from that, then what's the annotation preciseness you need}{Interviewee~F-II}
\noindent This inability to define success relates back to the inability of defining clear \emph{metrics} and KPIs for data and annotation quality.
\interviewquote{And I mean it's I think I can say they[, the OEMs,] push us [suppliers]. They want to have \textbf{more defined quality metrics} and so on, but it's also very hard for us to come up with them.}{Interviewee~C-II}
\noindent An interesting problems arises around the \emph{tools} used for the joint development of systems between suppliers and OEMs. Typically, in-house tools are proprietary and companies might be reluctant to share this intellectual property. A mentioned solution is the use and active development of open source tools:
\clearpage % 
\interviewquote{Yeah, so far we have not found any off the shelf product that can solve [what we need for collaboration], so \textbf{it's more or less a combination of in-house tools and open source tools}.}{Interviewee~C-I}
The ease of collaboration through open source tools might explain the success of such tools in the field of ML and data driven development, even in a competitive environment such as the automotive industry.
%
%%%%%%%%%%%%%%%%%%%%%%%%%%%%%%%%%%%%%%%%%%%%%%%%%%%%%%%%%%%%%%%%%%%%%%%%

\subsubsection{Shared responsibility}
New forms of partnerships between suppliers and OEMs causes shared responsibility for quality assurance, new \emph{business models} for data and annotation services, while still keeping due diligence of the OEM. This has some consequences on how responsibility is organised between the partners. \emph{Transparency} in the development processes becomes more important because building a safety case for perception software requires traceability and documentation of all design decisions, including traceability to the data and annotations used for the ML components:
\interviewquote{[...] talking from a safety case argumentation point of view; they will come, \textbf{they will ask for all the documentation and traceability} and they want to know what sort of process had you followed when it comes to machine learning as such. So we see a joint venture between us. To help out in the total vehicle certification point of view and when it comes to quality of the machine learning.}{Interviewee~D-I}
\noindent Building safety cases, and fulfilling the necessary quality criteria in both data and annotations is a \emph{collaborative} effort. Furthermore, close collaboration also enables OEMs better to reduce \emph{risk of litigation}, because in a collaborative environment, they better understand how the system was developed and what data has been used:
\interviewquote{We need to prove that we have done due diligence. So \textbf{it's not good enough to just believe what our supplier says that it will work}. [...] we need to to evaluate, how is that even possible and what data has been used and collect enough data in case there is a litigation and then we will also need to work with data, perhaps for a different business models.}{Interviewee~B-III}
%However, there are also unanswered legal question that arise in close collaborations: for example, who owns the data generated during the development in a joint project? The answer to this questions becomes significant when collaboration ventures beyond an OEM-supplier relationship and even includes the public through \emph{crowd-sourcing}. Crowd sourcing can mean, that customers, willingly or unknowingly, collect data for the companies which is then used to enhance the product.
%\interviewquote{No, I mean they [the OEM and the customers] will absolutely collaborate in ways, for example, by crowd-sourcing information about the world.}{Interview~C}

%%%%%%%%%%%%%%%%%%%%%%%%%%%%%%%%%%%%%%%%%%%%%%%%%%%%%%%%%%%%%%%%%%%%%%%%
%%%%%%%%%%%%%%%%%%%%%%%%%%%%%%%%%%%%%%%%%%%%%%%%%%%%%%%%%%%%%%%%%%%%%%%%
%%%%%%%%%%%%%%%%%%%%%%%%%%%%%%%%%%%%%%%%%%%%%%%%%%%%%%%%%%%%%%%%%%%%%%%%

\section{Discussion} \label{sec:discussion}
The results of the study gave an impression of the major change that is currently affecting the automotive sector: Data intensive developments, such as the development of software that includes ML models, cannot be conducted in the same way as ``traditional" software components. There is a lack of knowledge on how to properly specify aspects of data-intensive software. A notable finding of our study is that requirements specifications seem to play a major role in the sourcing and certification processes of automotive products, yet there is no common approach for specifying data or annotations of data. This is problematic in the sense that there are businesses emerging that specialise in the data procurement and data annotation for ML development, but automotive OEMs are not yet routinely able to collaborate with ``data" companies in the same manner as they can with system or components suppliers through established procurement processes and management.

\subsection{Recommendations}
%\eric{Consider using a table (theme, Recommendation, Comment, Researcg gap). Refer in the comment to literature. Otherwise, we fall short to situate our results in the state of the art.}
The interviews and the results from the validation workshop provided insight into the current state of practise when specifying data and annotations with the aim of achieving acceptable performance for automotive perception software that incorporates ML. From our observations, based on the lessons learned in the interviews, and the indicative scoring from the workshop, Table~\ref{tab:recommendation} provides several recommendations for practitioners in the industry.

\begin{table*}[]
\centering
\footnotesize
\caption{Recommendations for practitioners and towards researchers based on lessons-learnt in the Precog study}
\label{tab:recommendation}
\begin{tabular}{p{1.85cm}p{15.3cm}}
\hline
\multicolumn{1}{c}{\textbf{ID} \begin{scriptsize}(sub-category)\end{scriptsize}} &
  \multicolumn{1}{c}{\textbf{Recommendation}} \\ \hline
\textbf{Data-I} &
  \textbf{Establish clear traceability of data selection decisions as input towards a safety case of software with ML.} \\
 \begin{scriptsize}(D1-I, D2-II,\newline D3-I, B3-I)\end{scriptsize}&
  Comment: Data selection can have major implication on the correct behaviour of software that contains ML, because data selection strongly can influence bias and therefore should be traceable. 
 Related work: \cite{Hu2020, Gebru2021}
   \\ \hline
\rowcolor[HTML]{EFEFEF} 
\textbf{Data-II} &
  \textbf{Accept that a data specification can only be created iteratively.} \\
\rowcolor[HTML]{EFEFEF} 
 \begin{scriptsize}(D1-I, D2-I,\newline B3-II)\end{scriptsize}&
  Comment: Data preparations and selection for data-intensive software developments are obviously highly data driven activities. Practitioners need to analyse the available data before being able to understand which additional data might be needed. Therefore, conventional OEM-supplier sourcing processes might be unsuitable for data intensive developments and need to be changed. 
%\rowcolor[HTML]{EFEFEF} 
 Related work: \cite{Rahimi2019, Vogelsang2019}
   \\ \hline
\textbf{Data-III} &
  \textbf{Establish common metrics on data variation and other relevant data quality aspects to facilitate clearer communication between companies.} \\
 \begin{scriptsize}(D2-II, D2-III, D3\newline-I, D3-IV, D3-V)\end{scriptsize}&
  Comment: Data variation has been voted as the most important aspect of data quality, yet there is a lack of clear metrics that allow for specifying data variation. 
 Related work: \cite{Raji2020}  \\ \hline
\rowcolor[HTML]{EFEFEF} 
\textbf{Annotation-I} &
  \textbf{Evaluate if an increase in annotation quality in lieu of an increase of the annotation quantity, i.e., the amount of annotated data, can result in better ML model performance.} \\
\rowcolor[HTML]{EFEFEF} 
 \begin{scriptsize}(A1-I, A2-I,\newline A2-III)\end{scriptsize}&
  Comment: An increase in annotation quality seems to have stronger positive effects at the same cost compared to an increase in the amount of annotated data. 
%\rowcolor[HTML]{EFEFEF} 
 Related work: \cite{Marton2022, Schmarje2022}
   \\ \hline
   \textbf{Annotation-II} &
  \textbf{Concentrate on annotation consistency rather than pixel precision to increase annotation quality.} \\ 
 \begin{scriptsize}(A2-I, A2-II,\newline A2-VI, A2-VII)\end{scriptsize}&
  Comment: According to our interviewees, inconsistent annotations have worse effects on ML model performance than variations in the precision of the bounding boxes. 
 Related work: \cite{Tsipras2020, Taran2019}
   \\ \hline
\rowcolor[HTML]{EFEFEF} 
\textbf{Annotation-III} &
  \textbf{Clearly specify annotations and the annotation process.} \\
\rowcolor[HTML]{EFEFEF} 
 \begin{scriptsize}(A2-IV, A3-I, A3-II, B3-I, B3-IV)\end{scriptsize}&
  Comment: An annotation specification allows for judging the reliability of the annotated data, which can provide safety evidence and accountability towards the data-intensive software component. 
%\rowcolor[HTML]{EFEFEF} 
 Related work: \cite{Borg2019, Salay2018, Vaughan2017}
   \\ \hline
   \textbf{Ecosystems-I} &
  \textbf{Avoid conventional automotive OEM-supplier sourcing processes in data-intensive developments.} \\
 \begin{scriptsize}(B1-I, B2-II, \newline B2-III, B3-II)\end{scriptsize}& 
  Comment: An upfront requirements specification is often not feasible for data-intensive developments, because ``discovering" the right data and training a desired ML model is a highly iterative process. Instead, development partnerships with less bureaucracy are a trend mentioned by several interviewees. 
  Related work: \cite{Lempp2022}
   \\ \hline
\rowcolor[HTML]{EFEFEF} 
\textbf{Ecosystems-II} &
  \textbf{Increase the use of open source tools in shared data intensive developments.} \\
\rowcolor[HTML]{EFEFEF} 
 \begin{scriptsize}(B1-II, B2-I, \newline B3-I, B3-II,  \newline B3-III, B3-VI)\end{scriptsize}&
  Comment: Open source tools are from a legal perspective easier to share with new collaborators, they make it easier for smaller companies such as start-ups to participate in the development, and they establish transparency in the development process which can be positive when building a safety case or arguing for security. 
%\rowcolor[HTML]{EFEFEF} 
 Related work: \cite{Kochanthara2022}
   \\ \hline
\end{tabular}
\vspace{-1em}
\end{table*}

%%%%%%%%%%%%%%%%%%%%%%%%%%%%%%%%%%%%%%%%%%%%%%%%%%%%%%%%%%%%%%%%%%%%%%%%
%%%%%%%%%%%%%%%%%%%%%%%%%%%%%%%%%%%%%%%%%%%%%%%%%%%%%%%%%%%%%%%%%%%%%%%%
%%%%%%%%%%%%%%%%%%%%%%%%%%%%%%%%%%%%%%%%%%%%%%%%%%%%%%%%%%%%%%%%%%%%%%%%

\section{Threats to validity} \label{sec:threats}
Threats to validity can arise from the interviews, the workshop, and the data analysis process. 
In this section, we discuss possible threats to validity, and how we implemented mechanisms to reduce them.

%%%%%%%%%%%%%%%%%%%%%%%%%%%%%%%%%%%%%%%%%%%%%%%%%%%%%%%%%%%%%%%%%%%%%%%%

\subsection{Threats to internal validity}
Threats to internal validity arise when confounding variables cause bias in the result. This can occur through a lack of rigour (i.e., degree of control) in the study design \cite{Slack2001}. We established several mechanisms to reduce potential confounding: The interview guide was internally peer-reviewed and a test session of the interview was conducted before starting data collection. Furthermore, to avoid personal bias, at least two authors conducted each interviews. One of the authors was present at all interviews, while the others authors took turns in joining the interviews. After each interview, the authors aligned their interviewing experience in group meetings. The workshop was lead by all researchers, and a briefing after the workshop was conducted to share and discuss impressions obtained during the workshop. Another potential bias can arise from the sampling process. We deployed a mixture of purposeful and snowball sampling for both the interviews and the workshop. We needed a certain set of expertise to answer our questions, yet we also allowed companies to suggest additional interview partners. The companies were contacted through an open call. Additionally, we actively approached all OEMs in Sweden and received participants suggested by them. Furthermore, the workshop reduced potential selection bias, because participants outside of the companies of the interview study were included. Another threat to validity arises when saturation is not reached in the collected data. We can argue that we reached a point of saturation because we noticed a sharp decline in emerging codes after analysing the fifth group interview. 

%%%%%%%%%%%%%%%%%%%%%%%%%%%%%%%%%%%%%%%%%%%%%%%%%%%%%%%%%%%%%%%%%%%%%%%%

\subsection{Threats to external validity}
Threats to external validity arise when generalisability of the research results cannot be guaranteed. To support generalisability of the results, a sampling strategy was chosen that included different roles on different levels and at a number of different companies of different size. However, our study-results and conclusions are limited to the automotive sector, and specifically to the development of software for perception systems. However, we argue that perception system represent a typical situation in which a highly data-intensive software development is needed. Therefore, our results might also be valid for other data intensive development environments conducted in a more conservative business area like the automotive sector. An example can be the medical sector where ML plays more and more an important role in software for image based diagnostics.

%%%%%%%%%%%%%%%%%%%%%%%%%%%%%%%%%%%%%%%%%%%%%%%%%%%%%%%%%%%%%%%%%%%%%%%%
%%%%%%%%%%%%%%%%%%%%%%%%%%%%%%%%%%%%%%%%%%%%%%%%%%%%%%%%%%%%%%%%%%%%%%%%
%%%%%%%%%%%%%%%%%%%%%%%%%%%%%%%%%%%%%%%%%%%%%%%%%%%%%%%%%%%%%%%%%%%%%%%%

%%%%%%%%%%%%%%%%%%%%%%%%%%%%%%%%%%%%%%%%%%%%%%%%%%%%%%%%%%%%%%%%%%%%%%%%
%%%%%%%%%%%%%%%%%%%%%%%%%%%%%%%%%%%%%%%%%%%%%%%%%%%%%%%%%%%%%%%%%%%%%%%%
%%%%%%%%%%%%%%%%%%%%%%%%%%%%%%%%%%%%%%%%%%%%%%%%%%%%%%%%%%%%%%%%%%%%%%%%

\section{Conclusion and Outlook} \label{sec:conclusion}
This interview-based study investigated challenges that arise in the automotive industry when specifying data-intensive software components, such as software for perception systems. In seven group interviews with a total of 19 participants and through a validation workshop with 25 participants, we identified challenges that impact the ability to specify data and annotations of data. The inability to coherently measure data variation, unclear data collection processes, and the need of iterative development methodologies for data selection are examples of challenges that compromise the ability to specify data effectively for data depending software products in an automotive application. Unclear definition of annotation quality, a misleading focus on preciseness and quantity instead of consistency, and a lack of transparency in the annotation processes are examples of impediments that hinder proper annotation specifications. Furthermore, the study investigates current practises in the business environment and ecosystems deployed in the automotive industry, especially concerning a new trend towards emphasising joint development projects over the traditional OEM-supplier relationship in data-intensive developments. 
%We concluded this study by providing a first step towards an ontology ordering different aspects of data and annotations specifications into a causal order towards how they affect the performance of software components that contain trained ML algorithms. 
We concluded this study by providing a number of recommendations based on our observations. \par
We expect a major change in how the automotive industry is going to collaborate with suppliers and other partners in the development of data-intensive systems. The results of our study suggest a number of further research topics: The problem of defining clear metrics for data quality aspects and annotation aspects, and how partners can agree on proper metrics is not solved. There is research needed in understanding how different quality aspects of annotations should be specified for achieving a desired ML model performance. Furthermore, the development of an ``annotation industry" is in progress \cite{Sorokin2008}, and the success of these companies and ``crowd-sourcing" approaches depend on the ability to collaborate fruitfully with established companies such as OEMs in the automotive sector. Currently, an emphasis is set on quantity and precision over consistency in annotations. We need to learn how a suitable trade-off between different annotation aspects can be achieved, such that the cost for developing the software components is minimised and the resulting performance maximised.\par Based on the findings of this study, we propose further research 1) on how incremental collaborative specifications of data selection can be achieved, 2) on how such a specification for data selection can be validated, 3) on how the annotation process can be specified and eventually integrated in a safety assurance life-cycle, and 4) on how the information and knowledge sharing between OEMs and suppliers can be improved towards more joint responsibility in the development of machine learning models.
%%%%%%%%%%%%%%%%%%%%%%%%%%%%%%%%%%%%%%%%%%%%%%%%%%%%%%%%%%%%%%%%%%%%%%%%
%%%%%%%%%%%%%%%%%%%%%%%%%%%%%%%%%%%%%%%%%%%%%%%%%%%%%%%%%%%%%%%%%%%%%%%%
%%%%%%%%%%%%%%%%%%%%%%%%%%%%%%%%%%%%%%%%%%%%%%%%%%%%%%%%%%%%%%%%%%%%%%%%

\section*{Acknowledgements}
This project has received funding from Vinnova Sweden under the FFI program with grant agreement No 2021-02572 (precog), from the EU’s Horizon 2020 research and innovation program under grant agreement No 957197 (vedliot), and from a Swedish Research Council (VR) Project: Non-Functional Requirements for Machine Learning: Facilitating Continuous Quality Awareness (iNFoRM).
 We are thankful to all interviewees and companies who supported us in this research.

\bibliographystyle{IEEEtran}

\bibliography{bib}

%\begin{thebibliography}{00}

%\end{thebibliography}

\end{document}